\documentclass[aps,a4paper,eqsecnum,superscriptaddress,showpacs,pra
               ,preprint
              ]{revtex4}

\usepackage{amsfonts}
\usepackage{amsmath}
\usepackage{amscd}
\usepackage{amsthm}

\usepackage{epsfig}

\usepackage{bbm}

\newcommand{\e}{\ensuremath{\mathrm{e}}}
\newcommand{\Sym}{\operatorname{Sym}}  
\newcommand{\Dim}{\operatorname{dim}}
\newcommand{\NN}{\ensuremath{\mathbb{N}}}
\newcommand{\CC}{\ensuremath{\mathbb{C}}}
\def\id{{\rm 1\kern-.22em l}}

\DeclareMathOperator{\Det}{det}
\newcommand{\tr}{\operatorname{tr}} % Spur
\newcommand{\diag}{\operatorname{diag}}
\newcommand{\ad}{\ensuremath{a^\dagger}}

\newcommand{\ch}{\cosh}  % Cosinus-hyperbolicus
\newcommand{\sh}{\sinh}

\newcommand{\As}{\mathcal{A}}
\newcommand{\Bs}{\mathcal{B}}
\newcommand{\Cs}{\mathcal{C}}
\newcommand{\Ds}{\mathcal{D}}

\newcommand{\Ls}{\mathcal{L}} % Lax Operator

\newcommand{\xvec}{\ensuremath{\mathbf x}}

\newcommand{\XXX}{X\!X\!X}
\newcommand{\TQ}{\ensuremath{T}\ensuremath{Q}-}

\newcommand{\Matrix}[2]{\ensuremath{\left(\begin{array}{#1} #2 \end{array}}\right)}
\newcommand{\ket}[1]{\ensuremath{ | #1 \rangle}}
\newcommand{\bra}[1]{\ensuremath{ \langle #1 |}}
\newcommand{\braket}[2]{\ensuremath{ \langle #1 | #2 \rangle}}
\newcommand{\down}{\downarrow}
\newcommand{\up}{\uparrow}

\begin{document}

\title{Separation of variables for integrable spin-boson models}
%----------------------------------------------------------------------
\author{Luigi Amico\footnote[1]{e-mail: lamico@dmfci.unict.it}}
\affiliation{CNR-IMM MATIS $\&$ Dipartimento di Metodologie Fisiche e Chimiche
 (DMFCI), Universit\'a di Catania, viale A. Doria 6, I-95125 Catania, Italy}

\author{Holger Frahm\footnote[2]{e-mail: frahm@itp.uni-hannover.de}}
\affiliation{Institut f\"ur Theoretische Physik, Leibniz Universit\"at
 Hannover, Appelstr.\ 2, D-30167 Hannover, Germany}

\author{Andreas Osterloh\footnote[4]{e-mail: andreas.osterloh@uni-due.de}}

\affiliation{Fakult\"at f\"ur Physik, Universit\"at Duisburg-Essen, Campus
  Duisburg,\\ Lotharstr. 1, D-47048 Duisburg, Germany} 
\affiliation{Institut f\"ur Theoretische Physik, Leibniz Universit\"at
 Hannover, Appelstr.\ 2, D-30167 Hannover, Germany}

\author{Tobias Wirth\footnote[3]{e-mail:
    tobias.wirth@itp.uni-hannover.de}}
\affiliation{Institut f\"ur Theoretische Physik, Leibniz Universit\"at
 Hannover, Appelstr.\ 2, D-30167 Hannover, Germany}

%----------------------------------------------------------------------
\date{\today}

\begin{abstract}
  We formulate the functional Bethe ansatz for bosonic (infinite dimensional) representations
  of the Yang-Baxter algebra. The main deviation from the standard approach
  consists in a half infinite \emph{Sklyanin lattice} made of the eigenvalues of the operator zeros 
  of the Bethe annihilation operator.
  By a separation of variables, functional \TQ equations are obtained for this half infinite lattice.   
  They provide valuable information about the spectrum of a given Hamiltonian model. 
  We apply this procedure to integrable 
  spin-boson models subject to both twisted and open boundary conditions. 
  In the case of general twisted and
  certain open boundary conditions polynomial solutions to these \TQ equations are
  found and we compute the spectrum of both the full transfer matrix and
  its quasi-classical limit. For generic open boundaries we present
  a two-parameter family of Bethe equations, derived from \TQ equations that
  are compatible with polynomial solutions for Q. A connection of these parameters
  to the boundary fields is still missing.
  
 \end{abstract}

%02.30.-f Function theory, analysis
%05.30.Jp Boson systems 
%05.30.-d quantum statistical mechanics
%75.10.Pq Spin chain models
%02.30.Ik integrable Systems

\maketitle

%%%%%%%%%%%%%%%%%%%%%%%%%%%%%%%%%%%%%%%%%%%%%%%%%%%%%%%%%%%%%%%%%%%%%%
%
\section{Introduction}
%
%%%%%%%%%%%%%%%%%%%%%%%%%%%%%%%%%%%%%%%%%%%%%%%%%%%%%%%%%%%%%%%%%%%%%%
The method of separation of variables is a technique reducing a given  
multidimensional spectral problem to a set of uncoupled one dimensional 
equations. Although this reduction can in principle be applied to any 
integrable eigenvalue problem, its realization is a mathematically hard problem.
In this paper we focus on quantum integrable systems provided by 
the Quantum Inverse  Scattering Method (QISM). In this framework, 
representations of a quadratic Yang-Baxter algebra allow for the
construction of Hamiltonians from a family of commuting operators generated by
the transfer matrix. To actually solve the eigenvalue problem for these
operators Bethe ansatz methods are applied. For physical problems with 
a $U(1)$-symmetry the algebraic Bethe ansatz (ABA) \cite{KOREPIN-BOOK,Faddeev-notes} 
is the method of choice for this step. This approach and generalizations as 
presented in~\cite{RM,MRM} require the knowledge of a simple known eigenstate, 
the so-called reference state or pseudo vacuum. 
Unfortunately, it is the identification of this reference
state that is severely hampered in the absence of total-number conservation.
As a consequence, alternative methods are needed 
for the computation of the spectrum that avoid this difficulty.

Many of the existing alternatives, most notably Baxter's method of commuting
transfer matrices \cite{baxter:book} and Sklyanin's \emph{functional Bethe
  ansatz} (FBA) \cite{SklyaninNankai,SklyaninQuasi-Classical-FBA}, are based
on analytical properties of the model due to their construction within
the QISM and implicitly encode the eigenvalues into solutions to certain functional
relations. Various flavours of such functional methods have been successfully
applied to models where no reference state was known, including systems based
on non-compact symmetries (e.g. the quantum Toda chain~\cite{Sklyanin-Toda}
and the $\sinh$-Gordon model~\cite{Bytsko06,Teschner08}) or spin chains where
non-diagonal boundary fields break the $U(1)$-symmetry underlying the
applicability of the ABA
\cite{Nepo04,MRM,AmicoHikami05,MuNS06,YaNZ06,BaKo07,Galleas08,FrSeWi08}.

Unlike the ABA, this approach does not rely on the a priori knowledge of a
reference state.  Instead, the the representation of the Yang-Baxter algebra
underlying the integrable model is dealt with on a functional space isomorphic
to the Hilbert space of the model.  This allows to formulate the many-body
eigenvalue problem in such a way that it can be separated into equivalent
one dimensional equations.  This 'separation of variables'~\footnote{
  We want to mention that the separating variables of this approach are
  eigenvalues of mutually commuting but non-hermitean operators. Therefore,
  this separation is different from a standard separation of variables of a
  physical system, which is in terms of eigenvalues of observables, i.e. of
  hermitean operators.}
then leads to the functional (so-called \TQ) equations mentioned above.

In this work we apply the FBA to models that include a bosonic representation 
of the Yang-Baxter algebra for the case of ${\cal Y}[su(2)]$. 
Particular emphasise is given to integrable spin-boson models with  
a manifest violation of  the $U(1)$ symmetry; but also 
$U(1)$ symmetric model will be considered.

Interactions between spins and bosonic degrees of freedom are an archetypical
problem in many areas of physics.  They are encountered in the description of
systems as diverse as impurity atoms in condensed matter on top of a phononic
background, dissipative quantum systems, and of course all sorts of systems
involving the interaction of matter and radiation in quantum optics.  In
particular the class of systems, where atoms or ions are trapped and
controlled for various purposes~\cite{Lewenstein-Rev07,Bloch-Rev08}, has
experienced a boost of interest in the context of quantum information
processing and the simulations of complex physical systems.  Many of the
latter address problems from condensed matter physics, although there exist
also approaches to study quantum field theories this way (see e.g. Chapters
6\&7 of Ref.\cite{Lewenstein-Rev07}). The spin-boson interaction can be decomposed 
into what is called a \emph{rotating} part, i.e. $a S^+ + a^\dagger S^-$, and a corresponding
\emph{counter-rotating} part $a S^- + a^\dagger S^+$.  In the presence of only
one of these terms, say the rotating part, the interaction leaves the $U(1)$
charge $S^z+n_{\rm Bosons}$ conserved.  In this case, the Hamiltonian 
model is block-diagonal with separate blocks for each value of the conserved $U(1)$
charge~\cite{Jaynes63,Tavis68,Tavis69}.  
Many integrable extensions of these models have been found and analyzed
~\cite{Gerry88,Buzek89,Bogoliubov96,Rybin98}, where an integration 
of certain non-linear interaction terms has been achieved while 
keeping the $U(1)$ symmetry.
Hamiltonian models including both rotating \emph{and} counter-rotating terms 
have been obtained in Ref.~\cite{AmFrOsRi07} using the QISM, and by 
imposing suitable open boundary conditions.

The article is organized as follows: 
In the next section we sketch central elements of the QISM and give a concise
introduction to the functional Bethe ansatz for systems subject to
quasi-periodic (twisted) and open boundary conditions.  In
Section~\ref{FBA:bosons} we apply the FBA to models on an infinite dimensional
Hilbert space describing both spin and bosonic degrees of freedom.  Again, we
consider different boundary conditions: it is known that for arbitrary
quasi-periodic boundary conditions the spectrum can be obtained using the ABA.
Here we reconsider this case in an FBA approach in Section~\ref{twist}.
Open boundary conditions are relevant for the models with both rotating and
counter-rotating terms in the hamiltonian from Ref.~\cite{AmFrOsRi07}.  In
Section~\ref{open} we present in detail the derivation of the \TQ equations for
this case.  
The spectral problem for the spin-boson model with both types of boundary
conditions as encoded in these functional equations is investigated in
Section~\ref{Q-function}.  In either case we consider both the full transfer
matrix and its so-called \emph{quasi-classical limit}.  The procedure for
taking the latter is sketched in the appendix, it extends the family of
integrable hamiltonians which can be obtained within the approach used for the
construction of the model.  Within the FBA we find a complete solution of the
eigenvalue problem in terms of a set of algebraic 'Bethe' equations for
boundary conditions which can also be treated using the ABA.  Within the
generic set of open boundary conditions leading to both rotating and
counter-rotating interaction terms in the hamiltonian we propose a
three-parameter family of Bethe equations, which are obtained from the \TQ
equations but using a factorization of the quantum determinant that allows for
polynomial solutions for the function $Q$.

\section{Functional Bethe ansatz}

The functional Bethe ansatz method was originally formulated as a constructive way 
to realize a separation of
variables of a many body system, namely reducing a multidimensional problem to a 
suitable set of one dimensional ones\cite{SklyaninNankai}. The method relies on the concept of 
quantum integrability as provided by the
Quantum Inverse Scattering Method~\cite{KOREPIN-BOOK}. It may give insight 
to the exact spectrum for those systems where the ordinary algebraic Bethe ansatz
fails~\cite{Sklyanin-Toda}. In this section we sketch the functional Bethe ansatz for 
quantum integrable systems of interacting spins. 

The basic object of the QISM is the quantum $R$-matrix satisfying the Yang-Baxter equation
\begin{equation}\label{yangbaxter}
R_{12}(\lambda) R_{13}(\lambda+\lambda') R_{23}(\lambda') 
        = R_{23}(\lambda') R_{13}(\lambda+\lambda') R_{12}(\lambda)\;.
\end{equation}
It acts on a tensor product $V \otimes
V\otimes V$ of a vector space $V$ of a given dimension as
a function of the so-called spectral parameter $\lambda$.  
The indices of $R_{ij}$ indicate on which copy
of the tensor product the $R$-matrix acts non-trivially. 
The $R$-matrix can be exploited to provide the
commutation rules of an associative algebra  ${\cal T}$  
(quantum affine algebra) as
\begin{equation}\label{fundrel}
R_{12}(\lambda-\lambda') {\cal T}^{(1)}(\lambda) {\cal T}^{(2)}(\lambda') =
{\cal T}^{(2)}(\lambda')  {\cal T}^{(1)}(\lambda) R_{12}(\lambda- \lambda')\;,  
\end{equation}
where ${\cal T}^{(1)}={\cal T}\otimes \id$ and 
$ {\cal T}^{(2)}=\id \otimes {\cal T}$ and ${\cal T}$ can be
considered as an operator-valued matrix of dimension $\Dim V$. 
The vector space $V$ is referred to as the auxiliary space. 
The algebra ${\cal T}$ is of relevant interest in the theory 
of integrable quantum systems
because each of its representations provides a family of commuting operators. 
From this family a hamiltonian
model is deduced and the members of the family can then be considered as 
integrals of the motion.

In this paper, we will consider exclusively the rational solution of the Yang-Baxter equation 
identifying the Yangian affine algebra ${\cal Y}[sl(2)]$
\begin{equation} \label{Rmatrix}
R(\lambda,\mu) =
 \begin{pmatrix}
    1 & 0 & 0 & 0 \\
    0 & b(\lambda,\mu) & c(\lambda,\mu) & 0 \\
    0 & c(\lambda,\mu) & b(\lambda,\mu) & 0 \\
    0 & 0 & 0 & 1
 \end{pmatrix} \qquad , \qquad
 \begin{gathered}
    b(\lambda,\mu) = \frac{\lambda-\mu}
                          {\lambda-\mu+\eta} 
\\
    c(\lambda,\mu) = \frac{\eta}{\lambda-\mu+\eta}\end{gathered}\;.
\end{equation}

\subsection{Quasi-periodic boundary conditions}
Quantum models with periodic boundary conditions are constructed within QISM by 
choosing the representation of  
${\cal T}$ as
\begin{equation}
{\cal T}_{}(\lambda)={\cal L}_{L}(\lambda){\cal L}_{L-1}(\lambda)
\dots {\cal L}_{1}(\lambda)\;.
\label{monodromy}
\end{equation}
The matrix ${\cal L}_j$ is the so called Lax matrix. It is of dimension $\Dim V$ and has
operator-valued entries acting non-trivially in the quantum space of site $j$ only.
These Lax matrices also have to fulfill a Yang-Baxter equation    
\begin{equation}
R_{12}(\lambda-\lambda') {\cal L}_j^{(1)}(\lambda) {\cal L}_{j}^{(2)}(\lambda') =
{\cal L}_{j}^{(2)}(\lambda')  {\cal L}_{j}^{(1)}(\lambda) R_{12}(\lambda- \lambda'), \label{ybe-lax}
\end{equation}
with the additional requirement of ultra locality $\left [{\cal L}^{(1)}_{j}(\lambda),{\cal
L}^{(2)}_{l}(\lambda)\right ]=0\;$ for $j\neq l$.  Quasi-periodic boundary conditions 
can be used as the simplest
way to introduce boundary terms to the final hamiltonian. In the realm of the 
QISM this can be done slightly modifying ${\cal T}$ as
\begin{equation}\label{monodromy-twist}
  {\cal T}_{\text{twist}}^{(L)}(\lambda)=K {\cal T}(\lambda)\doteq
\begin{pmatrix}
  {A}(\lambda) & {B}(\lambda)\\ {C}(\lambda) & {D}(\lambda) 
\end{pmatrix}  \;.
\end{equation}
We comment that for the rational $R$-matrix from Eq.\eqref{Rmatrix}
the relation \eqref{fundrel} is satisfied by any $\lambda$ independent 
\CC-number matrix $K$ of dimension $\Dim V$ in ~\eqref{monodromy-twist}, 
since $[R_{12}(\lambda), K^{(1)} K^{(2)}]=0 $. 
The generating functional for the hamiltonian, and as well for the integrals of the motion, 
is the transfer matrix $t_{\text{twist}}(\lambda)=\tr_V {\cal T}_{\text{twist}}(\lambda)$ 
with the trace taken over the auxiliary space. 

The FBA method allows to construct separation of variables for the spectral problem
\begin{equation}
  t_{\text{twist}}(\lambda) \ket{\psi}=\Lambda(\lambda)\ket{\psi} \; .
\label{spectral_eq}
\end{equation}
The starting point of the procedure consists in looking at the operator valued zeros of the \lq lowering
operator\rq\  ${C}(\lambda)$ enjoying the property
\begin{equation} \label{C-commutation}
  [{C}(\lambda), {C}(\mu)]=0\; , \; \forall \lambda, \mu
\end{equation}
(equivalently, one can choose to consider operator zeros for the \lq raising operator\rq\ ${B}(\lambda)$ with  $[{B}(\lambda), {B}(\mu)]=0\; , \; \forall \lambda, \mu$).    
For $K_{21}\neq 0$, the meaning of the roots of the operator 
${C}(\lambda)$ can be specified by expressing the latter as \cite{SklyaninNankai} 
\begin{equation}
{C} (\lambda)=K_{21} \prod_{n=1}^L \left  (\lambda-\hat{x}_n \right )
\end{equation}
where the operators $\hat{x}_n$ can be simultaneously diagonalized because of the vanishing commutator
$[\hat{x}_n, \hat{x}_m]=0$ descending from the basic commutation relation (\ref{C-commutation}). 
In turn we observe that $ C(\lambda)$ can be diagonalized as it is indeed a polynomial 
operator of order $L$ in the spectral
parameter with coefficients that are symmetric functions of the roots $\hat{x}_1, \hat{x}_2,\dots,\hat{x}_L$.
Therefore the operators $ A(\lambda)$ and $ D(\lambda)$ are not diagonal in this basis.
We define
\begin{equation}
 \begin{aligned}
 A(\lambda=\hat{x}_n)&:=\sum_p \hat{x}_n^p {A}_p \equiv  \hat{X}^-_n\\
 D(\lambda=\hat{x}_n)&:=\sum_p \hat{x}_n^p {D}_p \equiv \hat{X}^+_n
\end{aligned}
\end{equation}
where an operator ordering is established by placing $\hat{x}$ to the very left in each term. 
We define $\Sym[\hat{x}_1,\dots, \hat{x}_L]$ as the set of symmetric functions of arguments $\hat{x}_1,\dots,\hat{x}_L$. 
The operators $X^\pm_n$ act on elements of $\Sym[\hat{x}_1,\dots, \hat{x}_L]$ as
\begin{equation} \label{X-sym}
\hat{X}^\pm_n \Sym[\hat{x}_1,\dots, \hat{x}_L] = \e^{\pm \eta \partial/\partial \hat{x}_n} \Sym[ (\hat{x}_1,\dots ,\hat{x}_L)]=\Sym[\hat{x}_1,\dots, \hat{x}_n \pm \eta, \dots, \hat{x}_L]\hat{X}^\pm_n
\end{equation}
suggesting them to be considered conjugated to the operators $\hat{x}_n$. In fact, 
the commutation rules are \cite{SklyaninNankai} 
\begin{equation} \label{fba-algebra}
  \begin{aligned}
    {}[\hat{X}^\pm_m, \hat{x}_n ]  & =\pm \eta  \hat{X}^\pm_m \delta_{mn} \\
{}[ \hat{X}^\pm_m, \hat{X}^\pm_n] &=  [\hat{X}^+_m, \hat{X}^-_n]=0\;. 
\end{aligned}
\end{equation}
On a generic (not necessarily symmetric) function $f(\hat{x}_1,\dots,\hat{x}_L)$ the operators $\hat{X}^\pm $ act as 
\begin{equation}
\hat{X}^\pm_n f(\hat{x}_1,\dots,\hat{x}_L)=\Delta^\pm(\hat{x}_n)  f(\hat{x}_1,\dots,\hat{x}_n\pm \eta, \dots,\hat{x}_L)
\label{X-extended}
\end{equation}
where $\Delta^\pm$ provide a factorization of the so called quantum determinant $\Det_q $ of the monodromy matrix  ${\cal T}_{\text{twist}}$.
The quantum determinant can be expressed as
$\Det_q ({\cal T}_{\text{twist}}) \equiv {A} (\lambda+\eta/2) {D}(\lambda-\eta/2)- {B} (\lambda+\eta/2){C}(\lambda-\eta/2)$ yielding directly
\begin{equation}
\hat{X}^\pm_n \hat{X}^\mp_n= 
\Delta^\pm (\hat{x}_n)\Delta^\mp(\hat{x}_n\pm\eta) =
\Det_q ({\cal T}_{\text{twist}} (\hat{x}_n\pm \eta/2)) \;.
\end{equation} 
Evaluating by substitution from the left the 
spectral equation \eqref{spectral_eq} at $\lambda=\hat{x}_n$ and sandwiching between left and right 
eigenvectors ${}_l\bra{x}$ and $\ket{x}_r$ of $ \hat{x}_1,\dots,  \hat{x}_n$ gives
\begin{equation}
\Lambda(x_n) \psi(x_1,\dots, x_L) =\Xi^+ \Delta^+(x_n)\psi( x_1,\dots,x_n+\eta,\dots,  \hat{x}_L)+
\Xi^- \Delta^-(x_n)\psi( x_1,\dots,x_n-\eta,\dots,  \hat{x}_L)
\end{equation}
where $\psi(x_1,\dots, x_L)\equiv {}_l\braket{x}{\psi}$ and 
      $\Xi^\pm = (\tr K\pm \sqrt{(\tr K)^2-4 \Det K})/2$. 
The final separation of variables is achieved by the ansatz 
$\psi(x_1,\dots, x_L)=\prod_j Q_j(x_j)$ leading to
\begin{equation}
\label{tq-spintwist}
\Lambda(x_n) Q_n(x_n) =\Xi^+ \Delta^+(x_n)Q_n( x_n+\eta)+
\Xi^- \Delta^-(x_n)Q_n( x_n-\eta)\;.
\end{equation}
These equations are $L$ {\it one dimensional} finite-difference equations for $x_n\in \mathfrak{G} $ with $\mathfrak{G}$ being the \lq lattice\rq\ provided by the eigenvalues of $\hat{x}_1,\dots, \hat{x}_n$.

\subsection{Open boundary conditions} 

The FBA method has been generalized recently to integrable models with 
open boundaries in Ref.{\cite{FrSeWi08}}.
In the seminal paper \cite{SklyaninOpen,CHOpen} Sklyanin demonstrated how to enlarge 
the class of integrable models obtainable from QISM by defining the so-called 
double-row transfer matrix describing a closed system interacting with a boundary. 
The core of the construction is the set of reflection algebras
\begin{equation} \label{eqref} 
  \begin{aligned} 
  &R_{12}(\lambda-\lambda') K^{(1)}_-(\lambda) R_{21}(\lambda+\lambda') K^{(2)}_-(\lambda') 
  = K^{(2)}_-(\lambda') R_{12}(\lambda+\lambda') K^{(1)}_-(\lambda) R_{21}(\lambda-\lambda')\;, \\ 
  &R_{21}(-\lambda+\lambda'){K^{(1)}_+}^t(\lambda) R_{12}(-\lambda-\lambda'-2\eta) 
    {K^{(2)}_+}^t(\lambda') \\
  &\qquad= {K^{(2)}_+}^t(\lambda') R_{21}(-\lambda-\lambda'-2\eta) 
       {K^{(1)}_+}^t(\lambda)R_{12}(-\lambda+\lambda'), 
\end{aligned} 
\end{equation} 
where $K(\lambda)$ parameterizes the boundary conditions, and $K^t$ is the transpose of $K$.
The involved $R$-matrix is again a solution of the Yang-Baxter equation \eqref{yangbaxter}.  
Additionally it fulfills the conditions of unitarity, parity, time reversal invariance, 
and crossing symmetry\cite{SklyaninOpen,CHOpen}.
It can be demonstrated that the following objects are representations of the 
reflection algebras \eqref{eqref}:
\begin{equation} 
  {\cal T}^{(+)}(\lambda)=K_+(\lambda)\quad , \quad {\cal T}^{(-)}(\lambda)= 
   {\cal T} (\lambda) K_-(\lambda) {\cal T}^{-1} (-\lambda)\;.  
\end{equation} 
Within this framework the generating functional for commuting integrals of the 
motion in Eq.\eqref{spectral_eq} is the following
operator (double row transfer matrix)~\cite{SklyaninOpen}
\begin{equation} 
  \label{doubletransf}
  t^{(L)}_{\text{open}}(\lambda)
  =\tr_V  {\cal T}^{(+)}(\lambda){\cal T}^{(-)}(\lambda)
  = \tr_V\left[ K_+(\lambda+\eta) {\cal T}(\lambda) K_-(\lambda)  {\cal T}^{-1}(-\lambda) \right]\; .  
\end{equation} 
As customary, we define $\mathcal{U}(\lambda) \equiv  \Det_q [{\cal T}(-\lambda-\eta/2)] {\cal T}^{(-)}(\lambda) $ with its matrix representation on the auxiliary space
\begin{equation} \label{U-matrix} 
  \mathcal{U}(\lambda)=
  \begin{pmatrix}
    {{\cal A}}(\lambda) & {{\cal B}}(\lambda)\\ 
    {{\cal  C}}(\lambda) & {{\cal D}}(\lambda)
  \end{pmatrix}
  \;. 
\end{equation} 
We particularly note that $[{{\cal B}}(\lambda),{{\cal B}}(\mu)]=0$ 
(for further relevant commutation relations between the operators in $\mathcal{U}(\lambda)$ 
see e.g.~\cite{SklyaninOpen}).  
In case of the rational $R$-matrix from Eq.\eqref{Rmatrix} the general non-diagonal 
\CC-number representations of the reflection algebras are 
the $K$-matrices~\cite{VeGo93} $K_+(\lambda)=\frac{1}{2}K(\lambda +\eta,+)$ and
 $K_-(\lambda)=K(\lambda,-)$ with
\begin{equation} \label{boundaries}
\begin{aligned} 
  K(\lambda,\pm) &= \frac{1}{\xi^\pm} 
  \begin{pmatrix} 
    \lambda + \xi^\pm & 2 \kappa^\pm \e^{\theta^\pm}\lambda \\ 
    2 \kappa^\pm \e^{-\theta^\pm}\lambda & - \lambda + \xi^\pm 
  \end{pmatrix} \\
  &\equiv \frac{1}{\alpha^\pm \ch \beta^\pm} 
  \begin{pmatrix} \lambda \sh \beta^\pm +\alpha^\pm
    \ch\beta^\pm & \lambda \e^{\theta^\pm} \\ 
	\lambda \e^{-\theta^\pm} & -\lambda \sh \beta^\pm +\alpha^\pm \ch\beta^\pm
  \end{pmatrix} \;, 
\end{aligned}
\end{equation} 
where (see Ref.~\cite{Nepo04}) $ \alpha^\pm \ch\beta^\pm = \frac{\xi^\pm}{2 \kappa^\pm}$ and 
$\sh\beta^\pm = \frac{1}{2\kappa^\pm}$.  
In the latter parametrization a diagonal boundary corresponds to the limit $\beta^\pm\to\infty$.

For the double-row transfer matrix \eqref{doubletransf} the FBA method proceeds 
through similar steps as discussed above for the quasi-periodic case but applied to the 
matrix $\mathcal{U}(\lambda)$ instead of ${\cal T}^{(L)}_{\text{twist}}(\lambda)$. 
${\cal B}(\lambda)$ in terms of its operator zeros can then be expressed as
\begin{equation} \label{operator-B}
  {\cal B}(\lambda)= (-)^L {\frac{2\lambda -\eta}{\alpha^-\ch\beta^-}} 
                     {\frac{\sh(\theta^- - \theta^+ -\beta^+)-\sh \beta^- }{2\ch\beta^+ } }
                      \prod_{l=1}^L (\lambda^2-\hat{x}_l^2)\;.
\end{equation}
The property $[\hat{x}_l^2,\hat{x}_m^2]=0$, arising from the commutation relations 
of the operators in $\mathcal{U}(\lambda)$, can be assumed as emerging from 
$[\hat{x}_l,\hat{x}_m]=0$. We sometimes write formally $\hat{x}_l =\diag\{x_l^+,x_l^-\} $ 
in terms of the eigenvalues $x_l^\pm$ of the non-hermitean operators $\hat{x}_l$.
The operators ${\tilde{{\cal D}}}(\lambda)\equiv 2 \lambda 
                {{\cal D}} (\lambda)-\eta {{\cal A}} (\lambda)$ 
and ${\tilde{{\cal C}}}(\lambda)\equiv (2\lambda-\eta) {{\cal C}}(\lambda)$ 
give rise to the shift operators $X^+_n={\cal A} (\hat{x}_n)$, 
and $X^-_n={\tilde{{\cal D}}} (\hat{x}_n)$. 
The action on generic functions is given in Eq.\eqref{X-extended} as the 
operator valued zeros provide the same algebra \eqref{fba-algebra} and provide the 
factorization of the quantum determinant 
$\Det_q \mathcal{U}(\lambda)={{\cal A}}(\lambda+\eta/2) \tilde{{{\cal D}}}(\lambda-\eta/2)
 - {{\cal B}}(\lambda+\eta/2) \tilde{{{\cal C}}}(\lambda-\eta/2)$:
\begin{equation}
  \begin{aligned}
X^-_nX^+_n 
&=\Delta^-(\hat{x}_n) \Delta^+(\hat{x}_n-\eta) ={\Det}_q U(\hat{x}_n- \eta/2)\\
X^+_nX^-_n 
&=\Delta^-(\hat{x}_n+\eta) \Delta^+(\hat{x}_n)= {\Det}_q U(\hat{x}_n+ \eta/2)
\end{aligned}
\end{equation}
where 
\begin{equation}
  \begin{aligned}
  \Delta^-(\lambda)&=(-)^L \frac{\lambda-\eta/2+\alpha^-}{\alpha^-}\prod_{l}(\lambda-x_l^-)(\lambda+x_l^+) \\
    \Delta^+(\lambda)&= (-)^L \frac{(\eta-2\lambda)(\lambda+\eta/2-\alpha^-)}{\alpha^-}\prod_{l}(\lambda-x_l^+)(\lambda+x_l^-) \;.
    \end{aligned}
\end{equation}

The spectral equation for $t^{(L)}_{\text{open}}(\lambda)$ from Eq.\eqref{doubletransf} reads as
\begin{eqnarray}
\Lambda(x_n) \psi(x_1,\dots, x_L) = \frac{(x_n+\eta/2)(x_n+\alpha^+-\eta/2)}{2 x_n \alpha^+}  \Delta^+(x_n)\psi( x_1,\dots,x_n+\eta,\dots,  \hat{x}_L)\\ +
\frac{x_n-\alpha^++\eta/2}{4 x_n \alpha^+}  \Delta^-(x_n)\psi( x_1,\dots,x_n-\eta,\dots,  \hat{x}_L)\; .
\nonumber
\end{eqnarray}
The ansatz $\psi(x_1,\dots, x_L)=\prod_j Q_j(x_j)$ again leads to an uncoupled set of 
one dimensional finite difference equations
\begin{equation}
\begin{aligned}
\Lambda(x_n) Q_n(x_n) =&\frac{(x_n+\eta/2)(x_n+\alpha^+-\eta/2)}{2 x_n
  \alpha^+} \Delta^+(x_n)Q_n( x_n+\eta)\\
 &+
\frac{x_n-\alpha^++\eta/2}{4 x_n \alpha^+}  \Delta^-(x_n)Q_n( x_n-\eta)\;.
\end{aligned}
\end{equation}
When we extend the validity of these \TQ equations to complex $x$,
only a single function $Q(x)$ remains. This will be our working hypothesis.

%%%%%%%%%%%%%%%%%%%%%%%%%%%%%%%%%%%%%%%%%%%%%%%%%%%%%%%%%%%%%%%%%%%%%%%%

%%%%%%%%%%%%%%%%%%%%%%%%%%%%%%%%%%%%%%%%%%%%%%%%%%%%%%%%%%%%%%%%%%%%%%%%

\section{Functional Bethe ansatz for spins interacting with a single bosonic mode} \label{FBA:bosons}
In this section we will carry out the functional Bethe ansatz for spin-boson hamiltonians. 
This corresponds to generalizing the approach to an infinite dimensional quantum space. 
In the framework of the QISM, integrable models for interacting bosons 
(e.g. photons) and spins (e.g. two-level atoms) have been constructed
from the algebraic structure induced by the $R$-matrix ~\eqref{Rmatrix}
for the rational six-vertex model~\cite{Bogoliubov96,AmicoHikami05,AmFrOsRi07}
using the boson and spin Lax operators
\begin{equation}
\label{laxe}
\Ls_b(\lambda) = 
\begin{pmatrix}
\lambda -\eta z_1 - \eta \ad a &  \beta \ad \\
\gamma a & -\frac{\beta \gamma}{\eta}
\end{pmatrix}; \quad
\Ls_s(\lambda) = 
\begin{pmatrix}
\lambda -\eta z_0 + \eta S^z &  \eta S^- \\
\eta S^+ & \lambda - \eta z_0 - \eta S^z\; ,
\end{pmatrix}
\end{equation}
where $z_0,z_1$ are inhomogeneities for the spin and the boson, respectively.
The quantum determinants are 
$\Det_q(\Ls_b)(\lambda) = -\frac{\beta \gamma}{\eta}(\lambda - (z_1-\frac{1}{2}) \eta)$ and 
$\Det_q(\Ls_s)(\lambda) = (\lambda -\eta z_0 -\eta)(\lambda  - \eta z_0 + \eta)$; 
the bulk monodromy matrix is defined as
\begin{equation}
\label{Mododromy:bulk}
{\cal T}(\lambda)\equiv\Ls_b(\lambda) \Ls_s(\lambda)\; .
\end{equation}

%%%%%%%%%%%%%%%%%%%%%%%%%%%%%%%%%%%%%%%%%%%%%%%%%%%%%%%%%%%%%%%%%%%%%
%
\subsection{Twisted boundary conditions}\label{twist}
The FBA for models with twisted boundary conditions
has been introduced in Ref.~\cite{SklyaninNankai} and applications include e.g.
the Gaudin model~\cite{SklyaninQuasi-Classical-FBA} and 
the Toda chain~\cite{Sklyanin-Toda}. Here we apply this technique
to models including a single bosonic degree of freedom.
An extension to more spins and/or bosons is straight forward.

Spin-boson models derived from the \XXX\ $R$-matrix defined in Eq.\eqref{Rmatrix} and with 
twisted boundary conditions are known to lead to models without 
counter-rotating terms~\cite{AmFrOsRi07}.
This is a consequence of the observation that every boundary twist matrix 
can be brought into upper triangular form by means of local gauge 
transformations~\cite{RM,MRM}, where the transformation matrices are
elements of the symmetry group of $R$~\cite{KulishSklyanin-YBE-Sols}. 
Although diagonalizable via the ABA, the application of the FBA to these simple models is 
still interesting for two reasons:
at first, we can compare with the results obtained from the usual
algebraic Bethe ansatz approach, and
second we will demonstrate explicitly how the FBA machinery works when
bosonic degrees of freedom are included.

We start from the monodromy matrix with general twist matrix $(K)_{jk}$
\begin{equation}
\label{twist_transf-spinboson}
{\cal T}^{(sb)}_{\rm twist}(\lambda) = \Matrix{cc}{K_{11} & K_{12}\\ K_{21}& K_{22}}\Ls_b(\lambda) \Ls_s(\lambda)
\end{equation}
and find that $ B$ has no term $\sim \lambda^2$. Instead, such a term is 
contained in $C$, and we will perform the FBA and separation of variables method
for the operator $C$ instead of $B$ (this ``asymmetry'' is a consequence
of the peculiar form of the bosonic Lax-operator).

It turns out to be convenient to consider linear combinations of $ B$ 
and $ C$ rather than $C$ directly. This can be realized by suitable similarity transformations 
of the monodromy matrix which do not affect the resulting transfer matrix.
We summarize this procedure in the monodromy matrix
\begin{equation}
\label{momodromy:twist}
{\cal T}^{(sb)}(\lambda) \equiv \Matrix{cc}{\alpha &\beta\\ 1& c^*}\Ls_b(\lambda) \Ls_s(\lambda)
\Matrix{cc}{1 & b\\ c & d}=
\Matrix{cc}{{A}&{B}\\ {C}& {D}}
\end{equation}
In complete analogy to Ref.~\onlinecite{SklyaninSepVars} we define the operator zeros
$\hat{x}_j$ of $ C$ as
\begin{equation}\label{C-zeros}
 C(\lambda)=\prod_{j=1}^N (\lambda-\hat{x}_j) = \sum_{j=0}^N (-1)^j\hat{c}_j \lambda^{N-j}\; .
\end{equation}
After subsequently applying the shift $a\rightarrow a+c$ to the bosonic 
operators for convenience, we find 
\begin{equation} \label{b-ops}
  \begin{aligned}
	\hat{c}_1&=-\eta\left[S_z+c S^- -\hat{n}\right]+z_0+z_1\\
	\hat{c}_2&=-\eta^2\left[\hat{n}(S_z+c S^-) 
           + \ad (2c S_z+c^2 S^- -S^+)\right]
	-\eta z_1\hat{c}_1 + (z_1-z_0) \hat{n} -z_0z_1\; .
\end{aligned}
\end{equation}
A common basis of right and left eigenstates of both operators is spanned by
\begin{equation}\label{rl-basis:twist}
  \begin{aligned}
    \ket{+,m}&=\ket{\down}\ket{m}+\frac{2\sqrt{m+1}}{2(m+z_1-z_0)+1}(\ket{\up}+c\ket{\down})\ket{m+1}\;
    \text{ for } m\geq 0 \\
    \ket{-,m}&=(\ket{\up}+c\ket{\down})\ket{m}\\
    \bra{+,m}&= -(c\bra{\up}-\bra{\down})\bra{m}\\
    \bra{r_{2,m}}&=\bra{\up}\bra{m}+\frac{2\sqrt{m}}{2(m+z_1-z_0)-1}(c \bra{\up}-\bra{\down})\bra{m-1}\;
    \text{ for } m \geq 0
  \end{aligned}
\end{equation}
and the corresponding eigenvalues defined by 
$\hat{c}_j\ket{\pm,m}=c_j^{\pm,m}\ket{\pm,m}$ are
\begin{equation}
c_1^{\pm,m}=\eta(m+z_0+z_1\pm\tfrac{1}{2})\quad ; \quad 
c_2^{\pm,m}=\eta^2(z_0\pm\tfrac{1}{2})(m+z_1)\;.
\end{equation}
The spin and boson operator zeros $\hat{x}_s$ and $\hat{x}_b$ are 
uniquely determined from the spectral decomposition as
\begin{equation}
  \begin{aligned}
\hat{x}_s&=-\eta\left[2\ad \frac{1}{2(\hat{n}+z_1-z_0)+1} (2c S_z+c^2 S^- -S^+)+c S^-+S_z-z_0\right]\\
\hat{x}_b&=\eta\left[\hat{n}+z_1+2\ad \frac{1}{2(\hat{n}+z_1-z_0)+1} 
(2c S_z+c^2 S^- -S^+)\right]
\end{aligned}
\end{equation}
and their eigenvalues are
\begin{equation}
x_{s,\pm}=\eta(z_0\pm\tfrac{1}{2})\quad ;\quad x_{b,m}=\eta(m+z_1)\; .
\end{equation}
In the limit of a diagonal boundary twist, pairs of the right and left eigenvectors of 
$\hat{x}_s$ and $\hat{x}_b$ coincide and the procedure breaks down.

Having the spectra of the operator zeros of $ C$,
we can explicitly write down a factorization of the quantum determinant
$\Det_q ({\cal T}^{(sb)}(\lambda))=\Delta^+(\lambda-\frac{\eta}{2})\Delta^-(\lambda+\frac{\eta}{2})$
that leads to \TQ equations after a separation of variables
\begin{equation}\label{sepvars}
Q(x_s,x_b)=Q(x_s)Q(x_b)\; .
\end{equation}
We find
\begin{equation}
\label{TQ-twist}
  \begin{aligned}
\Delta^+(\lambda)&= - \frac{\beta\gamma}{\eta} (\lambda-\eta(z_0+\tfrac{1}{2}))\\
\Delta^-(\lambda)&=(\lambda-\eta(z_0-\tfrac{1}{2}))(\lambda-\eta z_1)\\
\Lambda(x)Q(x)&=\Xi^+ \Delta^+(x)Q(x+\eta)+\Xi^-  \Delta^-(x)Q(x-\eta)
\end{aligned}
\end{equation}
with $\Xi^\pm$ from Eq.\eqref{tq-spintwist}.

\subsection{Open boundary conditions}\label{open} 
In order to create counter rotating spin-boson hamiltonians
we have to resort to open boundary conditions~\cite{SklyaninOpen,AmFrOsRi07}. 
In this case the FBA can be applied as well~\cite{FrSeWi08} as described above.

We consider the following transfer matrix
\begin{equation} \label{transfermatrix}
  t^{(sb)}_{\text{open}}(\lambda) = \tr K_+(\lambda - \eta/2) {\mathcal U}(\lambda)
\end{equation}
with ${\cal U}(\lambda)$ defined as
\begin{equation}
  \mathcal{U}(\lambda) \equiv {\cal T}(\lambda-\eta/2) K_-(\lambda-\eta/2) \sigma^y {\cal T}^t(-\lambda-\eta/2)\sigma^y \equiv \begin{pmatrix}
     \As (\lambda) & \Bs(\lambda) \\
     \Cs (\lambda) & \Ds(\lambda) 
   \end{pmatrix}\;.
\end{equation}
The boundary matrices are chosen as $K_-(\lambda)=K(\lambda,-)$ and $K_+(\lambda)=\tfrac{1}{2}K(\lambda+\eta,+)$ in terms of the $K$-matrix given in Eq.\eqref{boundaries}.

As $\Bs$ is a polynomial in the spectral parameter of maximal degree we directly factorize it in terms of its operator valued zeros. 
In Ref.~\onlinecite[Eq. (4.4)]{FrSeWi08} it was shown that
\begin{equation}
\Bs(\lambda) = - \frac{(2\lambda-\eta)\tanh \beta^-}{\alpha^-} \Bs_{\text{symm}}(\lambda)
\end{equation}
where $\Bs_{\text{symm}}(\lambda)= B_4 \lambda^4 + B_2 \lambda^2 + B_0$ is an even 
function of $\lambda$. 
An expansion of $\Bs_{\text{symm}}$ in its operator valued zeros $\hat x$ is 
\begin{equation}
\Bs_{\text{symm}}(\lambda) = \frac{\sh \beta^-- \sh(\theta^--\theta^+-\beta^+)}{2\sh \beta^-\ch\beta^+} (\lambda^2-\hat x_s^2) ( \lambda^2-\hat x_b^2) \; .
\end{equation}
We anticipate that the $\CC$-number zero $\lambda=\eta/2$ gives rise to an
additional constraint on the eigenvalue of $t^{(sb)}_{\text{open}}(\lambda)$ but 
will focus on the operator-valued zeros first.

Using the symmetric polynomials $b_1 = - B_2/B_4$ and $b_2 = B_0/B_4$ we obtain
\begin{equation}\label{b-x-relation}
\hat{b}_1 = \hat x_b^2 + \hat x_s^2 \quad; \quad  \hat{b}_2 = \hat x_b^2 \cdot \hat x_s^2
\end{equation}
The explicit coefficients $b_1$ and $b_2$ have been computed with FORM~\cite{FORM}.  

The bosonic Lax operator has entries proportional to a single $\ad$.
As $\ad$ does not have any right-eigenstates but $a$ does we will consider $b_1^\dagger$ and $b_2^\dagger$ in order to work with kets for convenience.
Expressed as a $2\times2$ matrix in spin space the operator $b_1$ reads 
\begin{equation}
\hat{b}_1^\dagger=\eta^2\Matrix{cc}{\hat{b}_1^{(++)}&\hat{b}_1^{(+-)}\\
                             \hat{b}_1^{(-+)}&\hat{b}_1^{(--)}} 
\end{equation}
with bosonic operator valued entries
\begin{align*}
\hat{b}_1^{(++)}&=-a^2 \frac{\beta^2}{\eta^2}\e^{-2\Theta^-}+ 
               a \frac{\e^{-\Theta^-} \beta}{\kappa\eta} \left(z_1 - \bar \xi+ n \right) 
                + z_0^2 +(z_1+ n+\frac{1}{2})^2 \\
\hat{b}_1^{(+-)}&= -2 a \e^{-2\Theta^-}\frac{\beta}{\eta}+\frac{\e^{-\Theta^-}}{\kappa} \left( z_0+\frac{1}{2} - \bar \xi \right) \\
\hat{b}_1^{(-+)}&=-2 a \frac{\beta}{\eta} \\
\hat{b}_1^{(--)}&=- a^2 \e^{-2\Theta^-} \frac{\beta^2}{\eta^2} + 
                   \frac{\e^{-\Theta^-}\beta}{\kappa\eta}
                           \left( z_1+2 - \bar \xi + n \right) a 
                       + (z_0+1)^2 + (z_1+ n+\frac{1}{2})^2 
\end{align*}
where $n=\ad a$ is the bosonic number operator and $\bar \xi \equiv \xi / \eta$.
The wave function is written correspondingly as a two-component vector
\begin{equation}
  \begin{aligned}
\ket{\psi} &= \sum_{n\sigma} \psi_{n\sigma} \ket{n\sigma}= \sum_{n=0}^{\infty} \left( \psi_{n\uparrow} \ket{n} \otimes \ket{\!\uparrow} + \psi_{n\downarrow} \ket{n}\otimes \ket{\!\downarrow} \right)\\
&= \sum_{n=0}^{\infty} \begin{pmatrix} \psi_{n\uparrow} \ket{n} \\ \psi_{n\downarrow} \ket{n} \end{pmatrix}\;.
\end{aligned}
\end{equation}
Considering the eigenvalue problem
\begin{math}
 \left(b_1^\dagger - E \id \right) \ket{\psi} = 0
\end{math},
where $E$ is the eigenvalue of $b_1^\dagger$, orthogonality 
leads to two intertwined recurrence relations for the coefficients $\psi_{n\sigma}$. 
Defining $\widetilde E \equiv E / \eta^2$,
$\delta_s \equiv z_0 + \tfrac12$, and $\delta_b \equiv z_1 + \tfrac12$ this implies
\begin{eqnarray}
0&=& -\sqrt{n+2}\sqrt{n+1} \e^{-2\theta^-}\psi_{n+2 \downarrow} + \sqrt{n+1} \frac{\e^{-\theta^-}}{\kappa} \left( (\delta_b-\frac{1}{2}+n-\bar \xi) \right) \psi_{n+1 \uparrow} +\\\nonumber 
&& \left( (\delta_s-\frac{1}{2})^2+(\delta_b+n)^2-\widetilde E \right) \psi_{n \uparrow} - 2 \sqrt{n+1} \e^{-2\theta^-}\psi_{n+1 \downarrow} + \frac{\e^{-\theta^-}}{\kappa} \left(\delta_s - \bar \xi \right) \psi_{n \downarrow} \\
0&=& -2 \sqrt{n+1} \psi_{n+1 \uparrow} - \sqrt{n+2}(n+1) \e^{-2\theta^-} \psi_{n+2 \downarrow} +
\sqrt{n+1} \frac{\e^{-\theta^-}}{\kappa} \times \\\nonumber
&& \left(\delta_b+\frac{1}{2}+n+\bar \xi \right)\psi_{n+1 \downarrow} + 
\left( (\delta_s+\frac{1}{2})^2+(\delta_b+n)^2-\widetilde E \right) \psi_{n\downarrow}
\end{eqnarray}
Considering the large $n$ regime ($n$ large as compared to the corresponding eigenvalues) each 
coefficient satisfies a $\Gamma$-function like functional relation.
Hence, in order to obtain normalizable states the recurrence relation must terminate 
at some finite boson number $m$,
and the eigenvalue can be read off directly from the coefficient of the 
highest boson number state $\ket{m}$ leading to
\begin{equation}
\begin{pmatrix}
\left( (\delta_b+m)^2+(\delta_s-\frac{1}{2})^2 - \widetilde E \right) \psi_{m \uparrow} + \left( \frac{\e^{-\theta^-}}{\kappa} (\delta_s - \bar \xi) \right)\psi_{m\downarrow} \\
\left( (\delta_b+m)^2 + (\delta_s+\frac{1}{2})^2 - \widetilde E \right) \psi_{m \downarrow} 
\end{pmatrix} \ket{m} = 0
\end{equation}
There are two possibilities for satisfying this set of equations
\begin{eqnarray}
\widetilde E^{b_1}_1&=& (\delta_b+m)^2+(\delta_s-\frac{1}{2})^2 \quad ; \quad \psi_{m\downarrow} = 0 \quad ; \quad \psi_{m\uparrow} \neq 0 \text{ arbitrary}\\
\widetilde E^{b_1}_2&=& (\delta_b+m)^2+(\delta_s+\frac{1}{2})^2 \quad ; \quad \frac{\psi_{m\uparrow}}{\psi_{m\downarrow}} = \frac{\e^{-\theta}}{2\kappa\delta_s}(\delta_s-\bar \xi)\quad.
\end{eqnarray}
The corresponding eigenstates can then be calculated by carrying out explicitly 
the recursion setting $\psi_{m+1,\scriptscriptstyle{\bullet}} = 0$ and 
$\psi_{m,\scriptscriptstyle{\bullet}}$ as stated just above.

An analogous calculation for the operator $b_2$ (with $\bar E \equiv E / \eta^4$) leads to
\begin{eqnarray}
\bar E^{b_2}_1&=& (\delta_b+m)^2 \cdot (\delta_s-\frac{1}{2})^2 \quad ; \quad \psi_{m\downarrow} = 0 \quad ; \quad \psi_{m\uparrow} \neq 0 \text{ arbitrary}\\
\bar E^{b_2}_2&=& (\delta_b+m)^2 \cdot (\delta_s+\frac{1}{2})^2 \quad ; \quad \frac{\psi_{m\uparrow}}{\psi_{m\downarrow}} = \frac{\e^{-\theta}}{2\kappa\delta_s}(\delta_s-\bar \xi) \quad.
\end{eqnarray}
The eigenstates again result from recurrence relations.

As $b_1$ and $b_2$ commute due to $[\Bs(\lambda),\Bs(\mu)]=0$ they share a common system of eigenvectors.
The eigenvalues of $b_1$ ($b_2$)  are only degenerate for a finite number of states if the inhomogeneities $z_0,z_1$ (resp. $\delta_{s/b}$) are chosen carefully. 
Typically, this degeneracy is lifted by the $b_2$ ($b_1$) operator. 
\footnote{For $z_0=0$ (i.e. $\delta_s=\frac{1}{2}$) and $z_1=-\frac{1}{2}$ (i.e. $\delta_b=0$) only the states for $m=0, \psi_{0\downarrow}\neq0,\psi_{0\uparrow}\neq0$ and $m=1,\psi_{1\uparrow}\neq0,\psi_{1\downarrow}=0$ are degenerate considering $b_1^\dagger$.} 
But e.g. the choice $z_0=-1/2$, i.e. $\delta_s=0$, results in a massive degeneracy 
of both $b_1$ \emph{and} $b_2$, and the separation of variables cannot be carried out 
in a straight forward way 
\footnote{Integer values for $z_1$ lead to a degeneracy if $2\delta_s$ is an odd positive integer;
the degeneracy then occurs for two consecutive integer boson quantum numbers $m=\delta_s+\frac{1}{2}$ 
and $\tilde{m}=m + 1$.}. 
It is a straight forward calculation that the above right-eigenstates of $b_1^\dagger$ are also
right-eigenstates of $b_2^\dagger$.

With the eigenvalues at hand it is possible to write the operator valued zeros of $\Bs$ as a matrix acting as multiplication operators on the common eigenbasis of $b_1^\dagger$ and $b_2^\dagger$ from Eq.\eqref{b-x-relation}.
For bosonic quantum number $m$ we find
\begin{equation}
\hat x_b^2 = \eta^2\begin{pmatrix} (\delta_b+m)^2 & 0 \\ 0& (\delta_b+m)^2 \end{pmatrix}
\quad ; \quad 
\hat x_s^2 = \eta^2\begin{pmatrix} (\delta_s-\frac{1}{2})^2 & 0 \\ 0& (\delta_s+\frac{1}{2})^2 \end{pmatrix}\;.
\end{equation}
As the $\hat x^2$ operators can be simultaneously diagonalized 
the operator zeros are 
\begin{equation}
\hat x_b = \eta\begin{pmatrix} \delta_b +m & 0 \\ 0& \delta_b+m \end{pmatrix}
\quad ; \quad 
\hat x_s = \eta\begin{pmatrix} \delta_s - \tfrac12 & 0 \\ 0& \delta_s + \tfrac12 \end{pmatrix}\;.
\end{equation}
This also fixes the sets $\mathfrak{G}$ for the lattice of the \TQ equations
\begin{eqnarray}
\mathfrak{G}_b &=& \{\delta_b\eta,(\delta_b+1)\eta,(\delta_b+2)\eta,\dots\} 
=:x_b^-+\NN_0 \; ; \\ 
\mathfrak{G}_s&=&\{\eta(\delta_s-\tfrac12),\eta(\delta_s+\tfrac12)\}
=:\{x_s^-,x_s^+\} \;.
\end{eqnarray}

As discussed above, the operator-valued zeros can be seen as \lq coordinates\rq\
whose \lq conjugated momenta\rq\ are the shift operators
\begin{equation}
\begin{aligned}
_{\lambda=\hat x_j}\left|\As(\lambda)\right. = \sum_p \hat{x}_j^p \As_p & \equiv X_j^-   \\
_{\lambda=\hat x_j}\left|\widetilde{\Ds}(\lambda)\right. = \sum_p
\hat{x}_j^p \widetilde{\Ds}_p& \equiv X_j^+ \quad .
\end{aligned}
\end{equation} 
where $\As_p$ and $\widetilde{\Ds}_p=2\lambda \Ds - \eta \As$ 
denote operator-valued expansion
coefficients of the operators of the algebra $\mathcal{U}$ and $j \in \{s,b\}$.
These \lq conjugated momenta\rq\ are representations of an algebra analog to
\eqref{fba-algebra}.
On arbitrary functions they act as 
$X_s^\pm f(\hat{x}_s,\hat{x}_b)=\Delta^\pm(\hat{x}_s)f(\hat{x}_s\pm\eta,\hat{x}_b)$,
$X_b^\pm f(\hat{x}_s,\hat{x}_b)=\Delta^\pm(\hat{x}_b)f(\hat{x}_s,\hat{x}_b\pm\eta)$
and induce a factorization of the quantum determinant of ${\cal U}$:
$\Delta^+(x-\eta/2)\Delta^-(x+\eta/2)=\Det_q{\cal U}(x)$ for all $x\in\mathfrak{G}$.
The following factorization meets these demands (compare \cite{FrSeWi08})
\begin{equation}
\begin{aligned}
\Delta^-(\lambda) &= \frac{\lambda-\eta/2+\alpha^-}{\alpha^-} 
(\lambda - \eta(\delta_s-\frac{1}{2}))(\lambda+\eta(\delta_s+\frac{1}{2}))( \frac{\beta\gamma}{\eta}( \lambda-\delta_b \eta)) \\
\Delta^+(\lambda) &= (2\lambda-\eta)\frac{\lambda+\eta/2-\alpha^-}{\alpha^-} 
(\lambda -\eta (\delta_s+\frac{1}{2}))(\lambda+ \eta(\delta_s-\frac{1}{2}))( \frac{\beta\gamma}{\eta}( \lambda+\delta_b \eta)) \;. 
\end{aligned}
\end{equation}
Note that $\Delta^\pm$ vanish on the appropriate boundaries of the sets $\mathfrak{G}_{s,b}$ 
especially the set $\mathfrak{G}_b$ is only bounded from below and hence only 
$\Delta^-(\delta_b \eta)$ is required to vanish. This is a consequence of the fact
that we have an infinite dimensional representation of the algebra~\eqref{fba-algebra}.
Given the lattice $\mathfrak{G}$ and the above factorization of $\Det_q{\cal U}$,
we arrive at the \TQ equations
\begin{equation}\label{TQ-equation:sb-open}
\begin{aligned}
\Lambda({x}_j) Q({x_j}) =& 
\frac{({x}_j+\eta/2)({x}_j+\alpha^+-\eta/2)}{2{x}_j\alpha^+}\Delta^-(x_j)Q(x_j - \eta) \\ 
&\qquad -\frac{{x}_j-\alpha^+ + \eta/2}{4{x}_j\alpha^+} \Delta^+(x_j) Q(x_j + \eta)  
\end{aligned}
\end{equation}
with an unknown function $Q(x)$.
The allowed arguments are $x_j \in \mathfrak{G}_j $ on the grid indentifying 
$\Delta_j^\pm(\xvec) = \Delta^\pm(x_j)$ for $j\in\{s,b\}$.

\section{The $Q$-function and the spectral problem}\label{Q-function}
In this section we investigate the possible information on the spectrum of the 
transfer matrix given by the \TQ equations arising from the FBA method. 
We remark, in this context, that the function $Q$ in its own does not carry any immediate
physical information. Instead, the values of $Q$ on the lattice $\mathfrak{G}$
might be the key to the eigenfunctions by means of a yet to be found
isomorphism. The basic working hypothesis consists in the 
identification of the latter equation with the Baxter \TQ equations, namely by 
extrapolating the validity of the \TQ equations to outside $\mathfrak{G}$
in interpreting them as a functional relation. 
We comment that the FBA method 
can be viewed at as being complementary to the ordinary algebraic Bethe ansatz 
for the spectral analysis (especially for those problems with no 
straightforward pseudo-vacuum states) 
in the sense that the method searches for a basis in which the 
'lowering operator' of the Yang-Baxter algebra is diagonal.

It is worth noticing that the \TQ equations derived on the lattice do not have a unique solution
in the space of continuous (or meromorphic) functions,
in particular not as far as the function $Q$ is concerned. 
Whereas the eigenvalue $\Lambda$ of the transfer matrix is known to be a polynomial with 
its degree given by the transfer matrix,
we do not have such a priori knowledge about the function $Q$:
additional knowledge about the latter is needed as an input. 

In those cases where a reference state can be found (possibly after gauge
transformations) and the algebraic Bethe ansatz can hence be performed,
polynomial solutions for $Q$ exist and lead to the known Bethe equations.
But even in these cases, no general direct link between $Q$ and the 
eigenfunctions is established in the literature.
A central open question concerns necessary and sufficient conditions
for a polynomial representative $Q$ to exist. 
We will not give an answer to this question, but will discuss specific cases,
which hopefully will shed some light onto this important problem.

\subsection{Quasi-periodic boundary conditions}\label{TQ:twist}

We start with the \TQ equations from \eqref{TQ-twist}.
We know from previous studies~\cite{AmFrOsRi07} that 
quasi-periodic boundary conditions always lead to algebraic Bethe ansatz solvable spin-boson models corresponding to slightly generalized Jaynes-Cummings hamiltonian (without counter-rotating terms).  According to the general scheme discussed above, we note  that the continuous limit of the FBA \TQ equations coincides with the Baxter equations found previously\cite{AmicoHikami05}. Therefore, each class of solutions for the \TQ equations
will have a polynomial representative:
$Q(\lambda)=\prod_{\alpha=1}^M(\lambda-\lambda_\alpha)$
leading to the Bethe equations
\begin{equation}
\label{BE-twist}
\frac{\eta K_{11}^2(\lambda_\beta-\eta z_0-\frac{\eta}{2})
                (\lambda_\beta-\eta z_1)}
    {\beta\gamma\Det K \cdot(\lambda_\beta-\eta z_0-\frac{\eta}{2})}=
\prod_{\alpha\neq \beta}\frac{\lambda_\beta-\lambda_\alpha+\eta}
                            {\lambda_\beta-\lambda_\alpha-\eta}
			  \end{equation}
and eigenvalues
\begin{equation}
\label{energy}
\Lambda(\lambda)=-\frac{\beta\gamma}{\eta K_{11}}\Det K 
     (\lambda-\eta z_0-\frac{\eta}{2}) 
                       \prod_\alpha\frac{\lambda-\lambda_\alpha+\eta}
                            {\lambda-\lambda_\alpha}
    +K_{11}(\lambda-\eta z_0-\frac{\eta}{2})
           (\lambda-\eta z_1)
                       \prod_\alpha\frac{\lambda-\lambda_\alpha-\eta}
                            {\lambda-\lambda_\alpha}\;.
\end{equation}
Both Bethe equations and eigenvalue equation agree with those in Refs.~\cite{Bogoliubov96,AmicoHikami05}.

\subsubsection{Quasi-classical expansion}
The quasi-classical expansion of the twisted spin-boson transfer 
matrix has been performed in Ref.~\onlinecite{AmicoHikami05} using
\begin{equation}
\label{amico-hikami-para}
K=\Matrix{cc}{-U-V+\sqrt{\frac{V}{U}} X \eta & X\eta \\ X \eta & 
-U-V+\sqrt{\frac{U}{V}}X \eta }
\;, \quad z_0 =0\;, \quad \gamma=\beta=1\;, \quad  z_1= \frac{1}{\eta^2}
\end{equation} 
leading to an integrable generalization of the Tavis-Cummings model; 
$X,U,V$ are real parameters; $U,V$ have the same sign. 
The parameters $Y$ and $\Delta$ in Ref.~\onlinecite{AmicoHikami05} have 
been set to zero here.
By inspection of the transfer matrix \eqref{twist_transf-spinboson} 
with the parameterization \eqref{amico-hikami-para}
the two lowest order contribution in powers of $\eta$ are proportional 
to the identity.
Therefore, the first non trivial term in the expansion 
(resulting $\propto \eta$) of the \TQ equations 
(see \ref{tq-quasiclassical_generic}) leads to the following Bethe equations
of the Gaudin type
\begin{equation}
\frac{1}{2}+\frac{2(U-1)\sqrt{V}+X\sqrt{U}}{2\sqrt{V}(V-U)} \lambda_k
-\lambda_k^2=\sum_{l \neq k} \frac{\lambda_k}{\lambda_k-\lambda_l}
\end{equation}
after the above polynomial ansatz for $Q$. 
These Bethe equations themselves determine a second order 
differential equation for $Q$:
\begin{equation}
\lambda Q''-   (1+\frac{X\sqrt{U}+2(U-1)\sqrt{V}}{\sqrt{V}(V-U)} 
\lambda+2\lambda^2)Q'(\lambda)+ (M \lambda -\zeta) Q(\lambda)=0
\end{equation}
where $\zeta=-\frac{Q'(0)}{Q(0)}=-\sum_\alpha \lambda_\alpha^{-1}$.

\subsection{Open boundary conditions}\label{TQ:open}

We consider the \TQ equations~\eqref{TQ-equation:sb-open} on the lattice $\mathfrak{G}$.
The vanishing of $\Delta^\pm$ on the boundaries of 
$\mathfrak{G}_s = \{ x_s^-\equiv \eta(\delta_s - \tfrac12), x_s^+ \equiv \eta(\delta_s + \tfrac12)\}$ yields the linear system of equations
\begin{equation}
\begin{aligned}
\Lambda(x_s^+) Q(x_s^+) =& \frac{(x_s^+ +\eta/2)(x_s^+ +\alpha^+-\eta/2)}{2x_s^+\alpha^+}\Delta^-(x_s^+)Q(x_s^-)\\ 
\Lambda(x_s^-) Q(x_s^-) =&  -\frac{x_s^--\alpha^+ + \eta/2}{4 x_s^- \alpha^+} \Delta^+(x_s^-) Q(x_s^+)  
\end{aligned}
\end{equation}
and $\Lambda(x_s^\pm)$ are obtained from the condition that
the determinant of the coefficient matrix of the linear system vanishes.
The vanishing of $\Delta^-$ at the \lq lower\rq\ boundary 
$\delta_b\eta\equiv x_b^0$ of 
$\mathfrak{G}_b = \{\delta_b \eta,(\delta_b+1)\eta,\dots,(\delta_b+n) \eta, \dots   \} 
\equiv \{x_b^0,x_b^1\dots, x_b^{n},\dots\} $ results in
\begin{equation}
\begin{aligned}
\Lambda(x_b^0) Q(x_b^0) =& \frac{(x_b^0 +\eta/2)(x_b^0 +\alpha^+-\eta/2)}{2x_b^0\alpha^+}\Delta^-(x_b^0)Q(x_b^{-1})
\end{aligned}
\end{equation}
and for the other lattice points we get ($n>0$)
\begin{equation}
\begin{aligned}
\Lambda(x_b^{n} ) Q(x_b^{n}) =& 
\frac{(x_b^{n}+\eta/2)(x_b^{n}+\alpha^+-\eta/2)}{2x_b^{n} \alpha^+}\Delta^-(x_b^{n})Q(x_b^{(n+1)}) \\ 
&\qquad -\frac{x_b^{n}-\alpha^+ + \eta/2}{4 x_b^{n} \alpha^+} \Delta^+(x_b^{n}) Q(x_b^{(n-1)}) \;.
\end{aligned}
\end{equation}
Here, $\Lambda(x_b^{n})$ are formally obtained from a {\em continuant} of a half-infinite 
matrix being zero. 
(A continuant is a determinant of a tridiagonal matrix: see e.g.~\cite{Muir-Dets}).

The expansion of the eigenvalue $\Lambda$ in powers of the spectral parameter 
$\lambda$ involves 4 coefficients. As the asymptotics is known, 3 equations are needed 
to fully determine $\Lambda$.
Besides the conditions of vanishing $2\times2$ determinant and half-infinite 
continuant we have a third equation from the \CC-number zero $\eta/2$ of the $\Bs$ operator. 
At this point the eigenvalue is the quantum determinant $\Det_q{\cal T}$ 
of the periodic monodromy matrix: $\Lambda(\eta/2) = \Det_q{\cal T}(-\eta/2)$. 
This can easily be seen by considering the unshifted and unscaled transfer matrix 
$\tilde{t}(\lambda) = \tr [ K_+(\lambda) {\cal T}(\lambda) K_-(\lambda) {\cal T}^{-1}(-\lambda) ]$.
Inserting $\lambda=0$ results in 
$\tilde{t}(0) = \tr [ K_+(0) {\cal T}(0) \id {\cal T}^{-1}(0) ] = \tr K_+(0) = 1$. 
The transfer matrices are related via 
$t(\lambda) = \Det_q{\cal T}(-\lambda) \tilde{t}(\lambda -\eta/2)$.

In the spirit of the scheme discussed in the beginning of the present section we now 
interpret the \TQ equations \eqref{TQ-equation:sb-open} 
as a functional relation which holds for general complex spectral parameter:
\begin{equation}\label{TQ2}
 \Lambda(\lambda) Q(\lambda) =
  \bar \Delta^- (\lambda)Q(\lambda-\eta)
 +\bar \Delta^+ (\lambda)Q(\lambda+\eta)\,.
\end{equation}
One finds that the coefficients 
\begin{equation}
\begin{aligned}
  \bar\Delta^- (\lambda) &=
 \frac{(\lambda+\eta/2)(\lambda+\alpha^+-\eta/2)}{2\lambda\alpha^+}\Delta^-(x_j)
 \\
       & = \frac{\beta\gamma}{2\eta\alpha^+\alpha^-}\,\frac{1}{\lambda}
           \left(\lambda+\frac{\eta}{2}\right)
           \left(\lambda+\alpha^+-\frac{\eta}{2}\right)
           \left(\lambda+\alpha^--\frac{\eta}{2}\right)\times\\
           &\qquad\times\left(\lambda+\eta(\delta_s+\frac{1}{2})\right)
           \left(\lambda-\eta(\delta_s-\frac{1}{2})\right)
           \left(\lambda-\eta\delta_b\right) \\
 \bar{\Delta}^+ (\lambda) &= \bar{\Delta}^- (-\lambda)
\end{aligned}
\end{equation}
behave asymptotically as $\lambda^5$ for large values of the spectral
parameter.  From the definition (\ref{transfermatrix}) of the transfer matrix
the eigenvalues for \emph{non-diagonal} boundary conditions
grow asymptotically as
\begin{equation}
\label{eigasy}
 \Lambda(\lambda) \propto
 \frac{ 2\e^{-\theta^+}  \e^{\theta^-}  }{\alpha^+\alpha^- \ch\beta^+
   \ch\beta^-} \lambda^6   + O(\lambda^4) \;.
\end{equation}
In the Bethe ansatz solvable case (diagonal or triangular boundary matrices)
the coefficient of $\lambda^6$ in (\ref{eigasy}) vanishes and the asymptotic
behaviour is $\propto \lambda^4$. Since in $\bar \Delta^\pm$ the leading order 
disappears as well, the \TQ equations \eqref{TQ2} can
by solved using an even polynomial ansatz for $Q(\lambda)$ in agreement
with the Bethe ansatz analysis in Ref.~\onlinecite{AmFrOsRi07}.

In the non-diagonal case the mismatch in the asymptotics for a chosen 
factorization of the quantum determinant must be compensated
by the $Q$-function. Hence, $Q$ will be transcendental. On the other hand,
we might want to insist on polynomial solutions for $Q$, 
but then the factorization 
must be modified accordingly \footnote{
It is clear that a polynomial ansatz for $Q$ requires that
the asymptotic limit for the spectral parameter in the \TQ equations
is governed solely by the eigenvalue polynomial $\Lambda(\lambda)$
and the quantum determinant factorization $\Delta^\pm(\lambda)$ 
(both polynomials for the \XXX\ model).}.

In order to allow for a polynomial form of $Q$, our intention is
to absorb a non-polynomial part in a function ${\cal F}$,
which in turn modifies the factorization of the quantum determinant. 
We hence define
\begin{equation} 
Q(\lambda)\equiv{\cal F}(\frac{\lambda}{\eta})\tilde{Q}(\lambda)
\label{modified-Q} 
\end{equation}
such that $\tilde{Q}(\lambda) $ is polynomial in $\lambda$ \footnote{Since every complex meromorphic function $M(z)$ can be written
uniquely as $M(z)=\e^{g(z)}\prod_j (z-z_{p,j})^{-1}\prod_k(z-z_{0,k})$
with an entire function $g(z)$, this is always possible (though not 
in a unique way).}.
To this end we assume that $\cal F$ satisfies the functional relation 
${\cal F}(z+1)=p(z) {\cal F}(z)$, where $p(z)$ is some rational function.
This leads to the following modified factorization of the quantum determinant
$\bar \Delta^\pm(\lambda)\longrightarrow 
\widetilde{\Delta}^\pm(\lambda)\equiv
\frac{{\cal F}(\frac{\lambda}{\eta}\pm 1)}{{\cal F}(\frac{\lambda}{\eta})}\bar \Delta^\pm(\lambda)
$ (representing an \lq algebra isomorphisms\rq\ as mentioned by Sklyanin
in Ref.~\onlinecite[Theorem 3.4]{SklyaninNankai}). On the other hand, 
this redefinition of $Q$ can be seen as a consequence of a change
in normalization for the right- and left eigenstates of $\mathcal{B}$,
which has no physical significance.
Therefore we can  redefine $Q$ such that
the asymptotics of the \TQ equations is fixed already by 
$\Lambda(\lambda)$ and $\tilde{\Delta}^\pm(\lambda)$, 
hence allowing for polynomial solutions $Q$.
The ansatz (\ref{modified-Q}) leads to the modified \TQ equations
\begin{eqnarray}
\Lambda(u)\tilde{Q}(u)&=&\bar{\Delta}^+(u)p(\frac{u}{\eta})\tilde{Q}(u+\eta)
+\bar{\Delta}^-(u)\frac{1}{p(\frac{u}{\eta}-1)}\tilde{Q}(u-\eta) \\
&=& \widetilde{\Delta}^+(u)\tilde{Q}(u+\eta)+\widetilde{\Delta}^-(u)\tilde{Q}(u-\eta)\;.
\label{TQ-tilde}
\end{eqnarray}
There are two possibilities for matching the asymptotics.
The first is to reach $\deg [\widetilde{\Delta}^-(u)]=\deg [\widetilde{\Delta}^+(u)]+2$ 
(\lq deg\rq\ is the polynomial degree).
Then
$p(u)\asymp u^{-1}$, and the ansatz
$p(u)=(p_\infty u\eta+\chi+p_\infty)^{-1}$ leads to
\begin{equation}
  \begin{aligned}
{\cal F}(z)=\Omega(z)\Gamma(-z-\zeta-1)\exp(-\alpha z)\\
\exp{\alpha}=- p_\infty \eta\; ;\ \zeta=\frac{\chi}{p_\infty \eta}
\end{aligned}
\end{equation}
where $\Omega(z)$ can be an arbitrary function with period $1$.
It does not affect the values on $\mathfrak G$.

In the above \TQ equations~\eqref{TQ2} we need 
$p_\infty=\frac{4\eta\e^{\theta^--\theta^+}}
             {\beta\gamma\ch \beta^+\ch \beta^-}$
and $\chi$ must be chosen suitably; one condition to be met would be
that no spurious upper bound for the bosonic spectrum is created.

The other possibility is $\deg [\widetilde{\Delta}^-(u)]=\deg [\widetilde{\Delta}^+(u)]-2$.
In this case, $p(u)\asymp u$, and the ansatz
$p(u)=(p_\infty u\eta+\chi)$ yields
\begin{equation}
  \begin{aligned}
\label{F-Gamma}
{\cal F}(z)=\Omega(z)\Gamma(z+\zeta)\exp(\alpha z)\\
\exp{\alpha}=p_\infty \eta\; ;\ \zeta=\frac{\chi}{p_\infty \eta}\; .
\end{aligned}
\end{equation}
The asymptotics of the \TQ equations ~\eqref{TQ2} is then fixed by the choice
$p_\infty=-\frac{4\eta\e^{\theta^- -\theta^+}}
             {\beta\gamma\ch \beta^+\ch \beta^-}$.

A formal polynomial ansatz $\tilde Q(\lambda)=\prod_{\alpha=1}^M(\lambda-\lambda_\alpha)(\lambda+\lambda_\alpha)$ 
leads to the Bethe equations
\begin{equation}
\frac{\tilde{\Delta}^-(\lambda_\beta)}{\tilde{\Delta}^+(-\lambda_\beta)
p(\frac{\lambda_\beta}{\eta})p(\frac{\lambda_\beta}{\eta}-1)}
=-
\prod_{\alpha\neq \beta}\frac{\lambda_\beta-\lambda_\alpha+\eta}
                            {\lambda_\beta-\lambda_\alpha-\eta}
\frac{\lambda_\beta+\lambda_\alpha+\eta}
                            {\lambda_\beta+\lambda_\alpha-\eta}
\end{equation}
in both cases.
Inspection of the transfer matrix suggests that the full degree $\lambda^6$ 
is present in the element corresponding to $\mathcal{D}$.
Assuming that this corresponds to the proper factorization for
non-diagonal boundaries, we insert the explicit linear form for $p(u)$
(the second case above). The Bethe equations derived from this 
assumption are
\begin{equation}
\label{BEqs:conjectured}
  \begin{split}
\frac{(\lambda_\beta+\frac{\eta}{2})(\lambda_\beta+\alpha^+-\frac{\eta}{2})
(\lambda_\beta+\alpha^--\frac{\eta}{2})(\lambda_\beta-\eta(\delta_s-\frac{1}{2}))
(\lambda_\beta+\eta(\delta_s+\frac{1}{2}))(\lambda_\beta-\eta\delta_b)
}{(\lambda_\beta-\frac{\eta}{2})(\lambda_\beta-\alpha^++\frac{\eta}{2})
(\lambda_\beta-\alpha^-+\frac{\eta}{2})(\lambda_\beta+\eta(\delta_s-\frac{1}{2}))
(\lambda_\beta-\eta(\delta_s+\frac{1}{2}))(\lambda_\beta+\eta\delta_b)
}\\
=-p_\infty^2(\lambda_\beta+\zeta\eta)(\lambda_\beta+(\zeta-1)\eta)
\prod_{\alpha\neq \beta}\frac{\lambda_\beta-\lambda_\alpha+\eta}
                            {\lambda_\beta-\lambda_\alpha-\eta}
\frac{\lambda_\beta+\lambda_\alpha+\eta}
                            {\lambda_\beta+\lambda_\alpha-\eta}
\; .
\end{split}
\end{equation}
It is evident that the parameter $\zeta$ plays a crucial role.
The case of diagonal boundaries is obtained in the limit $p_\infty\to 0$
and $\zeta\to\infty$ such that $p_\infty\zeta\to \chi/\eta$.
The known Bethe equations for the diagonal case~\cite{AmFrOsRi07,SklyaninOpen} 
are obtained if additionally $\chi\to 1$ in this limit.
It is easy to check that the functional relation for the function ${\cal F}$
reads ${\cal F}(z+1)={\cal F}(z)$ in this limit, which is satisfied by constant ${\cal F}$,
and hence no transformation of the \TQ equations is induced.
In case of linear $p(z)$, the simple zero $z_0=1-\chi/(p_\infty)$ of $p(z/\eta-1)$
must coincide with one of the zeros of 
$\tilde{\Delta}^-(z)\tilde{Q}(z-\eta)$~\footnote{In the analogous case that
$p(z)$ is a rational function with a single simple pole $z_p=-\chi/p_\infty$, 
this pole must coincide with a zero of $\tilde{\Delta}^+(z)\tilde{Q}(z+\eta)$.}.
In the \TQ equations this leads to the constraint 
\begin{equation}
-p_\infty\eta\Lambda(-\chi/p_\infty)\tilde{Q}(-\chi/p_\infty)=
\tilde{\Delta}^-(-\chi/p_\infty)\tilde{Q}(-\chi/p_\infty-\eta) \; .
\end{equation}

The correct $\chi$ will eventually be encoded in
the boundary matrices; it depends on the parameters
$p_\infty$ and $\zeta$. This is also seen from parameter counting:
besides the four eigenvalues, there remain two further parameters.
Therefore, the three parameters $p_\infty$, $\zeta$, and $\chi$
can not be independent. 

We emphasize that the factorization proposed above is 
only one possible choice and might not be
the correct one. In order to determine the proper factorization, 
a deeper understanding of the FBA is necessary.
We leave this open for future research.

%
%%%%%%%%%%%%%%%%%%%%%%%%%%%%%%%%%%%%%%%%%%%%%%%%%%%%%%%%%%%%%%%%%%%%%
%
\subsubsection{Quasi-classical expansion: diagonal $K$}
We first consider the \TQ equations \eqref{TQ2} for the Bethe ansatz solvable
case corresponding to diagonal boundary matrices (\ref{boundaries}), i.e.:
\begin{equation}
 K(\lambda) =  \frac{1}{\xi} 
 \begin{pmatrix} 
   \xi + \lambda & 0 \\
   0 & \xi - \lambda  
 \end{pmatrix}\,.
\end{equation}
For this choice the leading terms of the quasi-classical expansion of the transfer
matrix (\ref{transfermatrix}) in the parameter $\eta$ are
\begin{equation}
\begin{aligned}
 t^{(sb)}_{open}(\lambda) &= \frac{\beta\gamma}{\eta^2\xi^+\xi^-}
    \left(\tau^{(0)} + \tau^{(1)}\eta 
      + \tau^{(2)}\eta^2 + \ldots \right)\,,\\
 \tau^{(0)} &= 0\,,\qquad 
 \tau^{(1)} = (\xi^++\xi^-)\lambda^4\\
 \tau^{(2)} &= \left( 2S_z-2n -z_1-1\right) \lambda^4 
       + \left( \xi^+\xi^-(2S_z -z_1) + 2\xi^-\beta a^\dagger S^+ 
         \right) \lambda^2 \,.
 \end{aligned}
\end{equation}
The eigenvalues of $\tau^{(2)}$ in the sector with $n-S_z= (2k-1)/2$ are
\begin{equation}
\begin{aligned}
 \Lambda^{(2)}_{k,a}(\lambda) &=-\left( 2k+z_1\right) \lambda^4
       -\xi^+\xi^- \left(z_1+1\right) \lambda^2\,,
\\
 \Lambda^{(2)}_{k,b}(\lambda) &= -\left( 2k+z_1\right) \lambda^4
       -\xi^+\xi^- \left(z_1-1\right) \lambda^2\,.
\end{aligned}
\end{equation}
The $k=0$ subspace of the system is one dimensional, the corresponding
eigenvalue is $\Lambda^{(2)}_{k,a}(\lambda)$.
At second order in $\eta$  (\ref{TQ2}) turns into a first order 
differential equation for the eigenfunctions $Q(z)$, 
$z=\lambda^2$:
\begin{equation}
\begin{aligned}
 &z(z+\xi^+\xi^-)\, Q_{k,a}'(z) - (kz + \xi^+\xi^-)\, Q_{k,a}(z) =0\,,\\
 &(z+\xi^+\xi^-)\, Q_{k,b}'(z) -k\, Q_{k,b}(z) =0\,.
\end{aligned}
\end{equation}
Up to normalization these equations are solved by
\begin{equation}
\begin{aligned}
 &Q_{k,a}(z) \propto (z+\xi^+\xi^-)^k\,,\\
 &Q_{k,b}(z) \propto z (z+\xi^+\xi^-)^{k-1}\quad \mathrm{for~}k\ne0\,.
\end{aligned}
\end{equation}
As mentioned above, the $Q$-functions are even polynomials in the case of
diagonal boundary conditions -- this property still holds in the
quasi-classical limit.

Alternatively, we may solve the \TQ equations without using our knowledge of
the eigenvalues: with the ansatz $\Lambda^{(2)}(\lambda) = a_4\lambda^4 +
\xi^+\xi^-a_2 \lambda^2 + (\xi^+\xi^-)^2a_0$ we obtain the following
differential equation for $Q(z)$ (see Eq.~(\ref{tq-quasiclassical_generic})):
\begin{equation}
\begin{aligned}
 &2z^2\left(z+\xi^+\xi^- \right)Q'(z)\\
 &+ \left( \left(a_4+z_1\right) z^2 
         +\xi^+\xi^-\left(a_2 +z_1-1\right)z
         + (\xi^+\xi^-)^2 a_0\right) Q(z)=0\,.
\end{aligned}
\end{equation}
The general solution to this equation is
\begin{equation}
 Q(z) \propto {z}^{(1-a_2+a_0-z_1)/2} 
          \left( z+\xi^+\xi^- \right)^{(-1-a_4+a_2-a_0)/2}
          \exp\left(\frac{\xi^+\xi^-a_0}{2z}\right)\,.
\end{equation}
Requiring $Q(z)$ to be analytic with at most a simple zero at $z=0$ one
obtains immediately $a_0=0$ and the solutions given above is reproduced.

%%%%%%%%%%%%%%%%%%%%%%%%%%%%%%%%%%%%%%%%%%%%%%%%%%%%%%%%%%%%%%%%%%%%%%
\subsubsection{Quasi-classical expansion: non-diagonal $K$}
Generically, non-diagonal boundary matrices lead to non-hermitean
transfer matrices.  Within the quasi-classical approach, however, it is
possible to construct hermitean hamiltonians for the spin-boson model by fine
tuning the dependence of the system parameters on the 'quantum parameter'
$\eta$ \cite{AmFrOsRi07}: rescaling $z_0 \to z_0/\eta$ in Eq.~(\ref{laxe}) and
parametrizing the boundary matrices as
\begin{equation}
K(\lambda) =  \begin{pmatrix} \xi^\pm + \lambda & \lambda \mu^\pm \\
              \lambda \nu^\pm & \xi^\pm - \lambda  
 \end{pmatrix}
\end{equation}
the following choice of parameters
\begin{equation}
\label{smart1}
\begin{alignedat}{2}
 \mu^- &= \eta \mu_1^-\,, \quad 
 &\nu^- = \eta \nu_1^-\,, \quad
 &\xi^- = \eta\, \xi_1^-\,,\\
 \mu^+ &= \eta \frac{\beta}{\gamma} ( \mu^-_1+\nu^-_1)\,, \quad
 &\nu^+ = 0\phantom{ \nu_1^- }\,, \quad
 &\xi^+ = -\frac{\beta^2}{\eta}+\xi^+_0+ \eta \xi^+_1 \,.
\end{alignedat}
\end{equation}
leads to the following self-adjoint hamiltonian
\begin{equation}
\label{hamil-qc0}
 H=\Omega_0 n + \Delta_{\mathrm{sz}} S_z +\tfrac{1}{2}
\Delta_{\mathrm{sx}}(S^++S^-) 
 + g (S^+ \ad + S^- a)+2\alpha (a+\ad )
\end{equation}
where the coupling constants are obtained as
\begin{equation}
\begin{aligned}
& \Omega_0 = 2(z_0^2-\lambda^2), \quad
  \Delta_{\mathrm{sz}} = 2 (\lambda^2- \beta^2(\xi^-_1 -z_1)), \quad\\
& \Delta_{\mathrm{sx}} = -2 \beta^2 z_0 \nu^-_1, \quad
  g= 2 \beta z_0, \quad \alpha = \frac{\beta}{2}\nu_1^-(\lambda^2-z_0^2)
\end{aligned}
\end{equation}
in terms of the parameters in the transfer matrix.
Please note that $K^+$ has upper triangular form for this choice of parameters.
After a displacement of the bosonic operators $a \rightarrow a + \beta
\nu^-_1/2$ and a simultaneous rotation of the spin the hamiltonian becomes (up
to a constant) 
\begin{equation}
H=\Omega_0 n + \Delta_{\text{sz}} S_z + g (S^+ \ad + S^- a)\; .
\end{equation}
Although this operator has been obtained from non-diagonal boundary conditions
this operator commutes with the charge $n-S^z$. We could not find any 
choice of parameters leading to a hermitean hamiltonian including counter-rotating
terms. The same statement applies to a similar choice of parameters proposed in
Ref.~\onlinecite{AmFrOsRi07} -- different from what has been claimed there.
It is worth mentioning that within the above choice of $9$ parameters in the transfer matrix
only four (plus the spectral parameter $\lambda$) enter the final hamiltonian,
which can be obtained also from diagonal boundaries.
As some of these spurious parameters will be seen to enter the \TQ equations,
this is a hint that the function $Q$ indeed carries information about the
eigenstates of the hamiltonian, which are sensitive to local changes of the basis.   

It is furthermore intriguing that despite the presence of a conserved $U(1)$ charge
transformations such as \eqref{modified-Q} induced by a transcendental function
(see Eq.~\eqref{F-Gamma}) are required in order to guarantee solutions to the functional 
equations that can be parametrized by finitely many roots of a polynomial $\tilde{Q}(\lambda)$.
In contrast to the transformation of the factorization of the
quantum determinant of the full transfer matrix, 
a suitable $\eta$ dependence of the parameters (see above) is required
in the quasi-classical limit 
in order to ensure that the two lowest orders in $\eta$ remain
unchanged and that no lower orders are created by that transformation in the
\TQ equations. Taking account for these subtleties leads to the ansatz
\begin{equation}
  \begin{aligned}
Q(z)&={\cal F}(z)\tilde{Q}(z)\\
{\cal F}(z)&= (\eta^3\chi)^z \Gamma\left(z+\frac{1}{\omega\chi\eta^3}\right) \;.
\end{aligned}
\end{equation}
The quasi-classical expansion of the equation \eqref{TQ-tilde} is carried out
along the steps described in the appendix. 
The lowest orders $1/\eta^2$ and $1/\eta$ of the equation are identically 
satisfied for $\omega=1$; the order $\eta^0$ leads to the 
second order differential equation
\begin{equation}
\frac{1}{2}\Lambda_{{}_{-2}}(\lambda)\tilde{Q}'' (\lambda) 
  + R(\lambda) \tilde{Q}' (\lambda) 
  + U(\lambda) \tilde{Q}(\lambda) = \Lambda_0(\lambda) \tilde{Q}(\lambda) 
\label{fuchsian}
\end{equation}
which has the form of a Schr\"odinger equation for the Hamilton operator
$\mbox{${\cal H}=\frac{1}{2}\Lambda_{{}_{-2}}(\lambda)\partial_\lambda^2+R(\lambda)\partial_\lambda + U(\lambda)$}$.
\begin{equation}
  \begin{aligned}
\Lambda_{{}_{-2}}(\lambda)&=-\beta^3\gamma \lambda^2(\lambda^2-z_0^2)\\
R(\lambda)&\equiv -\frac{\beta\gamma}{2}\lambda\left[2\lambda^4-\lambda^2(2z_0^2+\beta^2(1-2z_1+2\xi^-_1))+\beta^2z_0^2(1+2z_1-2\xi^-_1)\right]\\
U(\lambda)&\equiv -\frac{\beta\gamma}{2}\left[\phantom{\frac{\beta^2}{2}}
\hspace*{-5mm}\lambda^4(\mu^-_1\nu^-_1\beta^2+2(z_1-\xi^-_1-\xi^+_1))\right. \\
&\qquad \qquad +\frac{\beta\gamma}{4}\lambda^2\left(-4z_0\xi^+_0+\beta^2(3-4\xi^-_1+4z_1(1+\xi^-_1))
\right.\\
& \left.\left. \qquad \qquad  +2z_0^2(2z_1+\mu^-_1\nu^-_1\beta^2+2(1-\xi^-_1-\xi^+_1))\right)
-\frac{\beta^3\gamma}{4} z_0^2(1+4z_1\xi^-_1) \right]\; .
\end{aligned}
\end{equation}
In the present case $\Lambda_0=\Lambda_0^{(0)}+\Lambda_0^{(1)}\lambda^2+\Lambda_0^{(2)}\lambda^4$. 
In particular, $\Lambda_0^{(0)}=U(\lambda=0)$,
which is consistent with a constant solution for $Q$, 
as should be expected for a rotating Jaynes-Cummings model.
$\Lambda_0$ is the energy as a function of the spectral parameter; it has to be 
fixed by the requirement that the corresponding differential equation \eqref{fuchsian}
has a polynomial solution, in analogy to the parameter 
$\zeta$ in Ref.~\onlinecite[Eq. (12)]{Bogoliubov96}.

It is seen that the spurious parameters $z_1$, $\mu^-_1$, $\nu^-_1$, $\xi^+_1$, 
and $\xi^+_0$ appear in the \TQ equations but not in the Hamiltonian.
$z_1$ can be safely set to zero, whereas $\mu^-_1$ and $\nu^-_1$ only 
occur in the invariant combination 
$\mu^-_1\nu^-_1\lambda^2={\xi^-_1}^2-\lambda^2-\frac{\det K^-}{\eta^2}$;
the same applies to $\xi^+$ in terms of the trace of $K^+$.
This indicates that their appearance in the \TQ equation reflects
a change in the eigenbasis encoded in the function $Q$.

It is worth noticing that the transformation leading to polynomial $\tilde{Q}$
in the quasi-classical limit does not fix the asymptotics of the \TQ equations
for the full transfer matrix. 
This highlights that the corresponding models differ considerably.

\section{Summary}
We have performed the separation of variables for spin-boson models 
generated from a bosonic Lax operator descending from rational 
six-vertex models with twisted and open boundary conditions. 
Our focus was on two-site compositions where a single spin interacts 
with a single bosonic mode.
Generic spin-boson interactions are counter-rotating and thus 
do not conserve $S^z+n$ in contrast to the rotating models. 
This hampers the construction
of a simple (pseudo) vacuum state, which is the necessary starting point for
a diagonalization of the model by means of the algebraic Bethe ansatz. 
We employed the functional Bethe ansatz, proposed by Sklyanin 
for systems without known reference vacuum state~\cite{SklyaninNankai}.
The application of this method to spin-boson models requires compatibility
of the formalism with an infinite dimensional bosonic Hilbert space. 
Our analysis demonstrates that the technique 
of separation of variables - originally designed for finite dimensional 
representation spaces - carries over straight forwardly to this scenario.
Specifically, we have found that the infinite dimensional bosonic 
Hilbert space is one-to-one reflected by a half-infinite Sklyanin lattice.
This provides further indication that the functional Bethe ansatz 
and its central features relying on the Yang-Baxter algebra alone do not depend
on the chosen representation.
The infinite dimensional representation of the 'factorization algebra' 
${\cal X}_\Delta$ (\onlinecite[Eq.(3.13)]{SklyaninNankai})
is due to the absence of a bosonic highest weight
and leads to a peculiar factorization of the quantum determinant,
which in turn is determined to some extent by
the boundary $\partial {\mathfrak G}=\{x_s^-,x_s^+\}\times\{x_b^-\}$
of the Sklyanin lattice ${\mathfrak G}$. 
As for finite dimensional representations (e.g. spin models), 
the Sklyanin lattice results to be composed of 
eigenvalues of the spin and boson part of the operator zeros of an 
off-diagonal monodromy matrix element~\footnote{Here we have chosen ${\cal C}$, which is
the vacuum annihilation operator in the algebraic Bethe ansatz solvable case.
Sklyanin used ${\cal B}$ in his proposal.}.
Interestingly enough, a suitable gauge transformation of the monodromy matrix
has facilitated the diagonalization of these operator zeros for the twisted model. 

The \TQ equations appear as a Schr\"odinger equation for the action of the 
transfer matrix on the Sklyanin lattice, where $Q$ is formally related to the eigenstates
of the model hamiltonian by some isomorphism~\cite{SklyaninNankai}. 
Algebraic Bethe equations are obtained from the \TQ equations for polynomial $Q$.
However, the function $Q$ is by no means uniquely defined.
A \TQ equation with a transformed function $Q$ can be obtained in general 
after transferring the effect of a transcendental part of $Q$ onto the factorization 
of the quantum determinant. 
In terms of the non-hermitean operator zeros $\hat{x}$
this induces a renormalization of their right (and left) eigenstates, which 
in turn is mathematically equivalent to the isomorphisms for the 
factorization algebra ${\cal X}_\Delta$ described in \cite{SklyaninNankai}. 
We have analyzed this ambiguity for the function $Q$ on a class of spin-boson hamiltonians 
with the aim to arrive at a \TQ equation allowing for a polynomial solution $Q$.
We find that the \lq canonical\rq\ factorization~\footnote{as given by the boundary 
of the Sklyanin lattice and fit to the known Bethe equations for
diagonal boundary matrices} 
of the quantum determinant of the transfer matrix will typically deviate from the
{\em proper} factorization~\footnote{i.e. a factorization that leads to polynomial solutions
for $Q$.} even in this simple case.

A presumable physical meaning of this {\em proper} 
normalization remains unclear and would be worth a further investigation. 
There are two pieces of evidence for a physical impact behind this choice 
of normalization. 
One is that different transformations are needed 
for the full transfer matrix and its quasi-classical limit. 
The second is that the Bethe equations 
are considerably changed in a way shown in Eq.~\eqref{BEqs:conjectured}. 
Understanding these issues would constitute a significant step
forward in the theory of integrable quantum systems.

%%%%%%%%%%%%%%%%%%%%%%%%%%%%%%%%%%%%%%%%%%%%%%%%%%%%%%%%%%%%%%%%%%%%%
\begin{acknowledgments}
We would like to thank S.~Niekamp and A.~Seel for valuable discussions. 
This work has been supported by the Deutsche Forschungsgemeinschaft under
grant number FR~737/6 and by the DAAD and CRUI sponsoring the German-Italian bilateral 
program VIGONI D/07/15337.
\end{acknowledgments}

\appendix
\section{Quasi-classical limit}
\label{qc-limit}

The so called quasi-classical limit of the
transfer matrix~\cite{SklyaninQuasi-Classical-FBA,HikamiKuWa-qclassical} consists in a series expansion in the 'quantum parameter' $\eta$ (playing the role of $\hbar$) of the transfer matrix around $\eta=0$:
$
\hat{t}(\lambda)= \hat{\tau}^{(0)} +
\eta\,\hat{\tau}^{(1)}(\lambda) + \eta^2 \,\hat{\tau}^{(2)} (\lambda) +\dots
$
with the aim of creating a commuting family of quasi-classical
transfer matrices $\tau^{(k)}(\lambda)$.
This procedure has proved to be particularly useful for extracting 'simple'
though non-locally interacting hamiltonians out of the transfer matrix.
Examples are the Gaudin magnets and corresponding 
BCS-like models.~\cite{vonDelft01,Amico01a,Amico01b,Sierra04PRB}

There is much freedom for introducing an $\eta$-dependence
to the boundary matrix parameters in integrable theories.
Examples for the quasi-classical limit of spin models with twisted 
and open boundary conditions can be found in Refs.~\onlinecite{HikamiGaudinopen,ADiLOHopen,ADiLOPRL}.

In the present cases (for twisted or open boundaries) the transfer matrix 
is a finite sum:
\begin{equation}
\hat{t}(\lambda)= \eta^{-k} \hat{\tau}^{(-k)}+ \dots + \hat{\tau}^{(0)} +
\eta\,\hat{\tau}^{(1)}(\lambda) + \eta^2 \,\hat{\tau}^{(2)} (\lambda) +\dots +\eta^m \,\hat{\tau}^{(m)} (\lambda)
\label{quasi-transfer}
\end{equation}
where $k$ and $m$ are integers.
Expanding the commutator relation $[\hat{t}(\lambda),\hat{t}(\lambda')]=0$ in $\eta$,
we obtain
\[[\hat{t}(\lambda),\hat{t}(\lambda')]=\sum_{l=-2k}^{2m} \eta^l C_l(\lambda,\lambda') = 0\;,\]
which implies $C_l(\lambda,\lambda')=0$ for all $l$.
The first relevant terms are
\begin{align}\label{commexp}
C_{-2k}(\lambda,\lambda')=&[\hat{\tau}^{(-k)}(\lambda),\hat{\tau}^{(-k)}(\lambda')]\;,\nonumber\\
C_{-2k+1}(\lambda,\lambda')=&[\hat{\tau}^{(-k)}(\lambda),\hat{\tau}^{(-k+1)}(\lambda')] +
[\hat{\tau}^{(-k+1)}(\lambda),\hat{\tau}^{(-k)}(\lambda')]\;, \nonumber \\
C_{-2k+2}(\lambda,\lambda')=&[\hat{\tau}^{(-k)}(\lambda),\hat{\tau}^{(-k+2)}(\lambda')]
+[\hat{\tau}^{(-k+2)}(\lambda),\hat{\tau}^{(-k)}(\lambda')]
+[\hat{\tau}^{(-k+1)}(\lambda),\hat{\tau}^{(-k+1)}(\lambda')]\;,\nonumber \\
\end{align}
From the expressions above, one finds that the first
$\hat{\tau}^{(n)}(\lambda)$ which is not a $\CC$-number (times the identity)
gives rise to a family of commuting operators.
Generically, the lowest order $\hat{\tau}^{(-k)}$ is a $\CC$-number. Therefore
the first class of integrable models are generated
by $[\hat{\tau}^{(-k+1)}(\lambda),\hat{\tau}^{(-k+1)}(\lambda')]=0$.
In the presence of boundary matrices, these are typically non-trivial operators
but representing non-interacting hamiltonians.
The task is then to tune the free parameters such that $\hat{\tau}^{(-k+1)}(\lambda)$
is also a $\CC$-number. 
The lowest non-trivial order in $\eta$,
e.g. $\hat{\tau}^{(-k+2)}(\lambda)$, is typically a hamiltonian with 
non-trivial interactions.

The \TQ equation (\ref{TQ2}) is a second order difference equation which has
to be solved to determine the spectrum of the spin-boson system.  In the
quasi-classical limit $\eta\to0$ it becomes a differential equation
\cite{SklyaninQuasi-Classical-FBA} for the $Q$-function we want to study here.
\begin{equation}
\label{tq-quasiclassical_generic}
\begin{aligned}
\sum_{j=-k}^m \eta^j \Lambda_j(\lambda) Q(\lambda) &= 
  \sum_{j=-k}^m \eta^j v^-_j(\lambda) \left (Q(\lambda)
       -\eta Q^{'} (\lambda)+\frac{\eta^2}{2} Q^{''}(\lambda) \right ) \\
&+ \sum_{j=-k}^m \eta^j v^+j(\lambda) \left (Q(\lambda)
       +\eta Q^{'} (\lambda)+\frac{\eta^2}{2} Q^{''}(\lambda) \right )
\end{aligned}
\end{equation}
where $v^\pm$ are suitable factorization of the quantum determinant (defined
above as $\Xi^\pm \Delta^\pm$ or $\tilde{\Delta}^\pm$ for quasi-periodic or
open boundaries respectively).  We point out that the differential equation
arising from the $n$-th order of the $\eta$-expansion involves (solely) the
coefficient $\Lambda_n$ of the eigenvalue; therefore, without the knowledge of
the latter, the differential equations arising from the \TQ equation can be
integrated only at a formal level.

%\bibliography{integrable,various,q-optics,BCS-Grains}  

%%%%%%%%%%%%%%%%%%%%%%%%%%%%%%%%%%%%%%%%%%%%%%%%%%%%%%%%%%%%%%%%%%%%%%%%

\end{document}